\documentclass[12pt,a4paper]{article}
\usepackage[usenames,dvipsnames]{xcolor}
\usepackage{jheppub}  
\usepackage{amssymb} 
\usepackage{amsmath}
\usepackage{amsthm}
\usepackage{mathtools}
\usepackage{amsfonts}    
\usepackage{dsfont}
\usepackage{pdfpages}
\usepackage{verbatim}
\usepackage{tensor}
\usepackage{mathrsfs}
\usepackage{siunitx}
\usepackage{url}

\usepackage{subcaption}
\usepackage{hyperref}
\usepackage{tikz}
\usetikzlibrary{patterns,decorations.pathreplacing, decorations.markings,calc,shapes.misc,decorations.pathmorphing}

\hypersetup{
    pdfencoding=unicode,
	colorlinks=true,
	urlcolor=Maroon,
	linkcolor=RoyalBlue,
	citecolor=Maroon,
	pdftitle={On the Virasoro fusion and modular kernels at any irrational central charge},
	pdfauthor={Julien Roussillon},
	pdfdisplaydoctitle=true,
	pdfstartview=FitH,
	linktocpage=true
}

\newcommand{\re}{\text{\upshape Re} \,}

 \newcommand\ben{\begin{equation*}}
 \newcommand\ebn{\end{equation*}}
 \newcommand\beq{\begin{equation}}
 \newcommand\eeq{\end{equation}}
 \newcommand\lb{\left(}
  \newcommand\rb{\right)} 

\newtheorem{cj}{Conjecture}

\let\Re\relax
\DeclareMathOperator\Re{Re}

\allowdisplaybreaks[1]

\title{On the Virasoro fusion and modular kernels at any irrational central charge}

\author{Julien Roussillon} 
\affiliation{Department of Mathematics and Systems Analysis \\ P.O. Box 11100, FI-00076, Aalto University, Finland.}
\emailAdd{julien.roussillon@aalto.fi}

\abstract{We propose a series representation for the Virasoro fusion and modular kernels at any irrational central charge.  Two distinct, yet closely related formulas are needed for the cases $c\in \mathbb C \backslash (-\infty,1]$ and $c <1$. Our proposal for $c <1$ agrees numerically with the fusion transformation of the four-point spherical conformal blocks, whereas our proposal for $c\in \mathbb C \backslash (-\infty,1]$ agrees numerically with Ponsot and Teschner's integral formula for the fusion kernel.  The case of the modular kernel is studied as a special case of the fusion kernel.}

\makeatletter
\gdef\@fpheader{}
\makeatother

\begin{document}

\maketitle

\section{Introduction and main results}

\subsection{Introduction}

Virasoro conformal blocks play a fundamental role in the conformal bootstrap approach to Conformal Field Theory in two dimensions. They are parts of correlation functions which are entirely determined by conformal symmetry. Consequently, they are special functions determined by representation theory of the Virasoro algebra. 

On a given Riemann surface, Virasoro conformal blocks form certain bases for the solution space of the Virasoro Ward identities \cite{Rib}. There exist linear transformations relating different bases of this space called \emph{crossing transformations}. In general, they are integral transformations whose kernels are called \emph{Virasoro crossing kernels}. A prototypical example is a special case where the crossing transformations on the four-point Riemann sphere reduce to the well-known connection formulas for the Gauss hypergeometric function \cite{HJP}.  

Of special importance is the case of the four-point Riemann sphere and of the one-point torus, because these two cases generate the set of all crossing kernels. The corresponding crossing kernels are denoted Virasoro fusion kernel and Virasoro modular kernels, respectively. Understanding the crossing properties of conformal blocks on these surfaces is primordial in the conformal bootstrap approach, since for instance the crossing symmetry equations for the correlation functions can be written in terms of the crossing kernels only \cite{Rib}. For a recent review of Virasoro conformal blocks, their crossing properties and connections to other areas of physics and mathematics, the reader is referred to \cite{E,Rib}. 

Let us now describe more precisely the different objects at play. We will use the following parametrization of the conformal dimensions and central charge:
\begin{align*}
    & \Delta(P) = \frac{Q^2}4 + P^2,  \qquad c = 1+6 Q^2,  \qquad Q = b+\frac{1}b.
\end{align*}
In the case $c \in \mathbb C \backslash (-\infty,1]$, the Virasoro fusion kernel $\mathbf F$ is defined by the following relation:
\begin{align}
 \label{fusionc>25} &  \mathcal F_{P_s}^{(b)}\left[\substack{P_2\;\;\;P_3\vspace{0.1cm}\\ P_1\;\;P_4}\right](z) = \int_{\mathbb R} dP_t \; \mathbf F^{(b)}_{P_s,P_t}\left[\substack{P_2\;\;\;P_3\vspace{0.1cm}\\ P_1\;\;P_4}\right] \mathcal F_{P_t}^{(b)}\left[\substack{P_2\;\;\;P_3\vspace{0.1cm}\\ P_1\;\;P_4}\right](1-z),
\end{align}
where $z$ is the cross-ratio of four points on the Riemann sphere, and where the conformal blocks $\mathcal F$ are defined in the natural normalization
\begin{equation}
\mathcal F_{P_s}^{(b)}\left[\substack{P_2\;\;\;P_3\vspace{0.1cm}\\ P_1\;\;P_4}\right](z) = z^{\Delta(P_s) - \Delta(P_1) - \Delta(P_2)} \lb 1 + O(z) \rb, \quad \text{as} \; z \to 0.
\end{equation}
The AGT correspondence \cite{AGT} provides an explicit power series representation in $z$ for $\mathcal F$. There also exist recursion relations for $\mathcal F$ due to Zamolodchikov \cite{Z1,Z2} which converge much faster than the AGT formula (see also \cite[section 2.4.2]{Rib}).

Throughout the paper we introduce the convention $s(a\pm b) = s(a+b)s(a-b)$. $\mathbf{F}$ admits the following integral formula due to Ponsot and Teschner \cite{PT1,PT2}: 
\begin{align} \nonumber
\mathbf F^{(b)}_{P_s,P_t}\left[\substack{P_2\;\;\;P_3\vspace{0.1cm}\\ P_1\;\;P_4}\right] = \; &\frac12  \frac{\Gamma_b(Q\pm 2iP_s)}{\Gamma_b(\pm 2iP_t)} \frac{\Gamma_b(\tfrac{Q}{2}-iP_2 \pm i P_3 \pm i P_t) \Gamma_b(\tfrac{Q}{2}+iP_4 \pm i P_1 \pm i P_t)}{\Gamma_b(\tfrac{Q}{2}-iP_2 \pm i P_1 \pm i P_s)\Gamma_b(\tfrac{Q}{2}+iP_4 \pm i P_3 \pm i P_s)} \\
\label{PTformula} & \times \int_{i\mathbb{R}} du \; \frac{S_b(\tfrac{Q}{4}-iP_2 \pm i P_1 + u) S_b(\tfrac{Q}{4}+iP_4 \pm i P_3 + u)}{S_b(\tfrac{3Q}{4}-iP_2 + iP_4 \pm i P_t + u) S_b(\tfrac{3Q}{4} \pm i P_s + u)}.
\end{align}
Here we use the standard notation for Barnes' double Gamma function $\Gamma_b(x)$, and the double Sine function $S_b(x)=\frac{\Gamma_b(x)}{\Gamma_b(Q-x)}$, see \cite[Appendix B]{E} for a review of their properties.

It is well known that the Ponsot-Teschner formula \eqref{PTformula} satisfies a set of shift equations in its momenta which originates from the pentagon relation \cite{E,R2021}. Eberhardt showed in \cite{E} that for $c \notin \mathbb Q$ and $c \in \mathbb C \backslash (-\infty,1]$, $\mathbf F$ is the unique solution of the shift equations that is meromorphic in all of its parameters. However, $\mathbf F$ does not admit an analytic continuation to $c \leq 1$, since the function $\Gamma_b$ is not defined in this region. Interestingly, Ribault and Tsiares discovered in \cite{RT} a transformation, the Virasoro-Wick rotation, which maps the unique meromorphic solution of the shift equations in the regime $c \in \mathbb C \backslash (-\infty,1]$ to the unique meromorphic solution for $c \in \mathbb C \backslash [25,\infty)$. However, the image of $\mathbf F$ under this transformation is an odd function of $P_s$ and $P_t$, hence it cannot be the physical fusion kernel for $c < 1$ since the conformal blocks are even. In view of Eberhardt's uniqueness result, Ribault and Tsiares were led to the conclusion that $\mathbf{\hat F}$ must have weaker analyticity properties than $\mathbf F$.

In what follows, we denote by $\mathbf{\hat F}$ the Virasoro fusion kernel for $c \leq 1$, or, equivalently, for $b \in i \mathbb R$. It will be convenient to describe it in terms of another set of parameters
\begin{equation} \label{complexification}
    \beta = ib, \qquad p = iP, \qquad \hat Q = \beta + \frac{1}{\beta}.
\end{equation}
Then, we define $\mathbf{\hat F}$ to be such that 
\begin{align}
 \label{fusionc<=1} &  \mathcal F_{P_s}^{(b)}\left[\substack{P_2\;\;\;P_3\vspace{0.1cm}\\ P_1\;\;P_4}\right](z) = \int_{i\mathbb R + \Lambda} dp_t \; \mathbf{\hat F}^{(\beta)}_{p_s,p_t}\left[\substack{p_2\;\;\;p_3\vspace{0.1cm}\\ p_1\;\;p_4}\right] \mathcal F_{P_t}^{(b)}\left[\substack{P_2\;\;\;P_3\vspace{0.1cm}\\ P_1\;\;P_4}\right](1-z).
 \end{align}
The choice of parameter dependence for $\mathbf{\hat F}$ will be justified in section \ref{derivation}. Note that in this case we are forced to shift the contour of integration by $\Lambda \in \mathbb R^*$ \cite{RS,RT} because the conformal blocks on the right-hand side have poles at
\begin{equation}
p_t^{(m,n)} = \frac{i}2(m\beta+n\beta^{-1}), \qquad m,n \in \mathbb N^*,
\end{equation}
hence the poles lie on the imaginary axis for $\beta \in \mathbb R$. Moreover, the result of the integral should not depend on $\Lambda$.
 
Finally, we define the Virasoro modular kernels $\mathbf M$ and $\mathbf{\hat M}$ to be the following special cases of $ \mathbf{F}$ and $\mathbf{\hat F}$ \cite{HJS}:
\begin{align} \label{relationFandM}
  &  \mathbf M^{(b)}_{P_s,P_t}[P_0] = \sqrt2 \; 256^{P_t^2 - P_s^2} \; \mathbf F^{(\sqrt 2 b)}_{\sqrt 2 P_s,\sqrt 2 P_t}\left[\substack{\frac{P_0}{\sqrt 2} \;\;\; \frac{ib}{2\sqrt 2} \vspace{0.1cm}\\ \frac{ib}{2\sqrt 2} \;\; \frac{ib}{2\sqrt 2}}\right], \\
  \label{relationFhatandMhat}  & \mathbf{\hat M}^{(\beta)}_{p_s,p_t}[p_0] = \sqrt2 \; 256^{-p_t^2 + p_s^2} \; \hat{\mathbf F}^{(\sqrt 2 \beta)}_{\sqrt 2 p_s,\sqrt 2 p_t}\left[\substack{\frac{p_0}{\sqrt 2} \;\;\; \frac{i\beta}{2\sqrt 2} \vspace{0.1cm}\\ \frac{i\beta}{2\sqrt 2} \;\; \frac{i\beta}{2\sqrt 2}}\right].
\end{align}
This originates from the fact that the one-point toric conformal blocks are special cases of the four-point spherical conformal blocks \cite{FLNO, Pog}.

\subsection{Main results}

In this paper, we propose a series representation for both $\mathbf F$ and $\mathbf{\hat F}$ as well as for the modular kernels $\mathbf M$ and $\mathbf{\hat M}$.  The idea that led to the proposals for $\mathbf F$ and $\mathbf{\hat F}$ is essentially to write a convenient ansatz for the series and solve the shift equations (see Section \ref{sec2p1} for more details). We emphasize that, by construction, the proposals for $\mathbf F$ and $\mathbf{\hat F}$ satisfy the same shift equations, however, the two are not related by analytic continuation. This is very analogous to the case of the structure constants of Liouville theory for $c>25$ and $c<1$ studied by Zamolodchikov in \cite{Z5}. Moreover, seeking series solutions of the shift equations was initiated by Nemkov in \cite{Nem} in the case of $\mathbf M$, building on earlier works \cite{GMM,Nem3} (see also \cite{Nem2} for ideas related to $\mathbf F$). It would be interesting to relate Nemkov's results for $\mathbf M$ to our proposal.

As a matter of convenience for the reader, all formulas are gathered below.

\subsubsection{The fusion kernels}

Our proposals for $\mathbf F$ and $\mathbf{\hat F}$ are as follows:
\begin{align} 
  \label{defF} & \mathbf F^{(b)}_{P_s,P_t}\left[\substack{P_2\;\;\;P_3\vspace{0.1cm}\\ P_1\;\;P_4}\right] = \frac12 \lb \mathbf F^{+,(b)}_{P_s,P_t}\left[\substack{P_2\;\;\;P_3\vspace{0.1cm}\\ P_1\;\;P_4}\right] + \mathbf F^{-,(b)}_{P_s,P_t}\left[\substack{P_2\;\;\;P_3\vspace{0.1cm}\\ P_1\;\;P_4}\right] \rb, \\
   \label{defhatF} & \mathbf{\hat F}^{(\beta)}_{p_s,p_t}\left[\substack{p_2\;\;\;p_3\vspace{0.1cm}\\ p_1\;\;p_4}\right] = \frac12 \lb\mathbf{\hat F}^{+,(\beta)}_{p_s,p_t}\left[\substack{p_2\;\;\;p_3\vspace{0.1cm}\\ p_1\;\;p_4}\right] + \mathbf{\hat F}^{-,(\beta)}_{p_s,p_t}\left[\substack{p_2\;\;\;p_3\vspace{0.1cm}\\ p_1\;\;p_4}\right]\rb,
\end{align}
where $\mathbf F^{-,(b)}_{P_s,P_t} = \mathbf F^{+,(b)}_{-P_s,P_t}$, $\mathbf{\hat F}^{-,(\beta)}_{p_s,p_t} = \mathbf{\hat F}^{+,(\beta)}_{-p_s,p_t}$, and where $\mathbf F^+$ and $\hat{\mathbf F}^+$ take the form
\begin{align}
  \label{defF+}  & \mathbf F^{+,(b)}_{P_s,P_t}\left[\substack{P_2\;\;\;P_3\vspace{0.1cm}\\ P_1\;\;P_4}\right] = K_{P_s,P_t}^{(b)} \left[\substack{P_2\;\;\;P_3\vspace{0.1cm}\\ P_1\;\;P_4}\right] f_{P_s,P_t}^{(b)} \left[\substack{P_2\;\;\;P_3\vspace{0.1cm}\\ P_1\;\;P_4}\right], \\
   \label{defhatF+} & \mathbf{\hat F}^{+,(\beta)}_{p_s,p_t}\left[\substack{p_2\;\;\;p_3\vspace{0.1cm}\\ p_1\;\;p_4}\right] = \hat K_{p_s,p_t}^{(\beta)} \left[\substack{p_2\;\;\;p_3\vspace{0.1cm}\\ p_1\;\;p_4}\right] f_{p_s,p_t}^{(\beta)} \left[\substack{p_2\;\;\;p_3\vspace{0.1cm}\\ p_1\;\;p_4}\right].
\end{align}
The factors $K$ and $\hat K$ are given by
\begin{align} \label{definitionKc>1}
    & K_{P_s,P_t}^{(b)} \left[\substack{P_2\;\;\;P_3\vspace{0.1cm}\\ P_1\;\;P_4}\right] = e^{i\pi \lb P_1^2+P_2^2+P_3^2+P_4^2+\frac{1+b^2+b^{-2}}{4}\rb} \\
  \nonumber  & \times \frac{\Gamma_b(2i P_s)\Gamma_b(Q+2i P_s)}{\Gamma_b(-2i P_t)\Gamma_b(Q-2i P_t)} \frac{\Gamma_b(\frac{Q}2-iP_t \pm i P_2 \pm i P_3)\Gamma_b(\frac{Q}2-iP_t \pm i P_1 \pm i P_4)}{\Gamma_b(\frac{Q}2+iP_s \pm i P_2 \pm i P_1)\Gamma_b(\frac{Q}2+iP_s \pm i P_3 \pm i P_4)}, \\
  \label{definitionKc<=1} & \hat K_{p_s,p_t}^{(\beta)} \left[\substack{p_2\;\;\;p_3\vspace{0.1cm}\\ p_1\;\;p_4}\right] = -i e^{i\pi \lb p_1^2+p_2^2+p_3^2+p_4^2+\frac{1+\beta^2+\beta^{-2}}{4}\rb} \\
  \nonumber & \times \frac{\Gamma_\beta(2i p_t+\frac{1}{\beta})\Gamma_\beta(2i p_t+\beta)}{\Gamma_\beta(-2i p_s + \frac{1}{\beta})\Gamma_\beta(-2i p_s + \beta)} \frac{\Gamma_\beta(\frac{\hat Q}2-ip_s \pm i p_2 \pm i p_1)\Gamma_\beta(\frac{\hat Q}2-ip_s \pm i p_3 \pm i p_4)}{\Gamma_\beta(\frac{\hat Q}2+ip_t \pm i p_2 \pm i p_3)\Gamma_\beta(\frac{\hat Q}2+ip_t \pm i p_1 \pm i p_4)}.
\end{align}
It remains to describe $f$ which is an infinite series of the form
\begin{equation} \label{ansatz}
    f_{P_s,P_t}^{(b)} \left[\substack{P_2\;\;\;P_3\vspace{0.1cm}\\ P_1\;\;P_4}\right] := e^{2i\pi P_s P_t} \lb\sum_{k=0}^\infty \alpha_k(b,P_s) e^{-2\pi b k P_t}\rb \lb\sum_{l=0}^\infty  \alpha_l(\tfrac{1}b,P_s)  e^{- \frac{2\pi l P_t}b}\rb.
\end{equation}
We provide a recursive representation for the coefficients $\alpha_n$:
    \begin{align}
     \label{alphan}  &  \alpha_n(b,P_s) = \delta_{n,0} \\
 \nonumber & + \sum_{l=1}^n e^{-\pi b(P_s + \frac{lib}2)} \left[\frac{\phi_{ln}(P_1,P_2,P_3,P_4)}{\sinh(\pi b(P_s + \frac{lib}2))} - \frac{\phi_{ln}(P_1,P_2+\frac{i}{2b},P_3+\frac{i}{2b},P_4)}{\cosh(\pi b(P_s + \frac{lib}2))}\right].
    \end{align}
The coefficients $\phi_{nn}$, $n>0$ have the explicit form
    \begin{align}
  \label{phinn} & \phi_{nn}(P_1,P_2,P_3,P_4) = \frac{i 2^{2n-1} (-1)^{n+1}}{\sin(n\pi b^2)\prod_{l=1}^{n-1} \sin(l\pi b^2)^2} \\ 
        \nonumber & \times \prod_{l=1}^n \cosh\left(\pi b\big(\pm P_1 + P_2 + ib\lb l-\tfrac{n+1}2\rb\big)\right) \cosh \left(\pi b\lb P_3 \pm P_4 +  ib(l-\tfrac{n+1}2\rb\big)\right),
    \end{align}
whereas the coefficients $\phi_{ln}$ for $0< l \leq n$ have the semi-explicit form
    \begin{equation} \label{philn}
        \phi_{ln}(P_1,P_2,P_3,P_4) = \phi_{ll}(P_1,P_2,P_3,P_4) \; \alpha_{n-l}\lb b, \frac{lib}2 \rb.
    \end{equation}
For instance, $\alpha_1$ explicitly reads 
\begin{align}
\label{alpha1} & \alpha_1(b,P_s) = \frac{2i e^{-\pi b(P_s+\frac{ib}2)}}{\sin(\pi b^2)} \\
\nonumber & \times \lb \frac{\cosh(\pi b(P_2 \pm P_1)) \cosh(\pi b(P_3 \pm P_4))}{\sinh(\pi b(P_s + \frac{ib}2))} - \frac{\sinh(\pi b(P_2 \pm P_1)) \sinh(\pi b(P_3 \pm P_4))}{\cosh(\pi b(P_s + \frac{ib}2))} \rb.
\end{align}
More generally, utilizing \eqref{alphan} inductively, $\phi_{ln}$ can be expressed only in terms of $\phi_{mm}(P_1,P_2,P_3,P_4)$ and $\phi_{mm}(P_1,P_2+\frac{i}{2b},P_3+\frac{i}{2b},P_4)$ for $m=1,...,n-l$.

\subsubsection{The modular kernels}

We now proceed with the case of the modular kernels. We verified numerically that the formulas below for $\mathbf M$ and $\mathbf{\hat M}$ satisfy \eqref{relationFandM} and \eqref{relationFhatandMhat}, respectively. We have
\begin{align} 
  \label{defM}  & \mathbf M^{(b)}_{P_s,P_t}[P_0] = \frac12 \lb \mathbf M^{+,(b)}_{P_s,P_t}[P_0] + \mathbf M^{-,(b)}_{P_s,P_t}[P_0] \rb, \\
   \label{defhatM} & \mathbf{\hat M}^{(\beta)}_{p_s,p_t}[p_0] = \frac12 \lb \mathbf{\hat M}^{+,(\beta)}_{p_s,p_t}[p_0] + \mathbf{\hat M}^{-,(\beta)}_{p_s,p_t}[p_0]\rb,
\end{align}
where $\mathbf M^{-,(b)}_{P_s,P_t} = \mathbf M^{+,(b)}_{-P_s,P_t}$, $\mathbf{\hat M}^{-,(\beta)}_{p_s,p_t} = \mathbf{\hat M}^{+,(\beta)}_{-p_s,p_t}$, and
\begin{align}
  \label{defM+}  & \mathbf M^{+,(b)}_{P_s,P_t}[p_0] = L_{P_s,P_t}^{(b)}[P_0] \; g_{P_s,P_t}^{(b)}[P_0], \\
   \label{defhatM+} & \mathbf{\hat M}^{+,(\beta)}_{p_s,p_t}[p_0] = \hat L_{p_s,p_t}^{(\beta)}[p_0] \; g_{p_s,p_t}^{(\beta)}[p_0].
\end{align}
The factors $L$ and $\hat L$ read
\begin{align}
    & L_{P_s,P_t}^{(b)}[P_0] = \sqrt 2 \; e^{\frac{i\pi}2 (\frac{Q^2}4+P_0^2)} \frac{\Gamma_b(2iP_s)\Gamma_b(Q+2iP_s)}{\Gamma_b(-2iP_t)\Gamma_b(Q-2iP_t)} \frac{\Gamma_b(\frac{Q}2 \pm  iP_0-2iP_t)}{\Gamma_b(\frac{Q}2 \pm iP_0+ 2iP_s)}, \\
    & \hat L_{p_s,p_t}^{(\beta)}[p_0] = -i \sqrt 2 e^{\frac{i\pi}2 (\frac{\hat Q^2}4+p_0^2)} \frac{\Gamma_\beta(2ip_t+\frac{1}{\beta})\Gamma_\beta(2ip_t+\beta)}{\Gamma_\beta(-2ip_s+\frac{1}{\beta})\Gamma_\beta(-2ip_s+\beta)} \frac{\Gamma_\beta(\frac{\hat Q}2 \pm i p_0 - 2ip_s)}{\Gamma_\beta(\frac{\hat Q}2 \pm i p_0 + 2ip_t)}.
\end{align}
Finally, the series is of the form
\begin{equation} \label{seriesg}
    g_{P_s,P_t}^{(b)}[P_0] = e^{4i\pi P_s P_t} \sum_{k=0}^\infty \sum_{l=0}^\infty \mu_k(b,P_s) \mu_l(\tfrac{1}b,P_s) e^{-4\pi b k P_t} e^{-\frac{4\pi l P_t}b},
\end{equation}
and the coefficients $\mu_k$ have an explicit form in terms of $q$-Pochhammer symbols. More precisely, let 
\begin{equation} \label{qpoch}
(a,q)_n = \prod_{k=0}^{n-1} (1-a q^k).
\end{equation}
and denote $\alpha_0 := \frac{Q}2+iP_0$. Then, we have
\begin{align} \label{mun}
   & \mu_n(b,P_s) \\
 \nonumber   & = \sum_{l=0}^n \frac{\lb e^{2i\pi b \alpha_0}; e^{2i\pi b^2} \rb_l}{\lb e^{2i\pi b^2} ; e^{2i\pi b^2} \rb_l} \frac{\lb e^{-2\pi b(2P_s + i\alpha_0)} ; e^{-2i\pi b^2} \rb_l}{\lb e^{-2\pi b(2P_s + ib)} ; e^{-2i\pi b^2} \rb_l} \frac{\lb e^{-2i\pi b \alpha_0} ; e^{2i\pi b^2} \rb_{n-l}}{\lb e^{2i\pi b^2} ; e^{2i\pi b^2} \rb_{n-l}} e^{2i\pi b \alpha_0(n-l)}.
\end{align} 

\subsection{Numerical tests}

The details of the numerical tests described in this section can be found in the ancillary Jupyter notebook, or in the GitLab repository \cite{Ropython}.

\subsubsection{The fusion transformation for $c<1$}

We performed a test of the fusion transformation \eqref{fusionc<=1} with the following values of the parameters:
\begin{equation} \label{valueparam} \begin{pmatrix}
\beta \\ p_1 \\ p_2 \\ p_3 \\ p_4 
\end{pmatrix} = \begin{pmatrix}
 0.6 \\  0.4i \\  0.6i \\ 0.3 i \\  0.5i
\end{pmatrix}, \qquad p_s = 0.35i, \quad z = 0.5, \quad \Lambda = 0.5.
\end{equation}
We used the publicly available GitLab repository \cite{Ribpython} developed by Ribault to compute values of the function $\Gamma_b$ and of the conformal blocks. 

The accuracy of the computations is controlled by three parameters $(T,L,N_\text{max})$.\footnote{To reach an accuracy of $\sim 10^{-16}$ we also decreased slighlty the value of the parameters epsabs and epsrel of the Python method quad. We do not mention it in the main text for convenience.} The parameter $T$ represents the truncation of the two series in f \eqref{ansatz} at order $T$. $L$ represents the truncation of the infinite integration line $i \mathbb R + \Lambda$ to $i[-L,L]+\Lambda$. Finally, $N_\text{max}$ corresponds to the truncation of Zamolodchikov's recursion relation for the conformal blocks \cite{Ribpython}. At the values \eqref{valueparam}, the left-hand side of \eqref{fusionc<=1} is real-valued and equals approximatively 1.08. Then, we made the following verifications:

\begin{center}
\begin{tabular}{|c|c|c|c|} 
\hline 
  (T,  $N_\text{max}$, L) & $\left|\text{l.h.s - r.h.s of \eqref{fusionc<=1}}\right|$ \\ \hline
  (0,10,1) & $1.2$ \\ \hline
  (1,12,2) & $0.3$ \\ \hline
  (2,14,3) & $0.01$ \\ \hline
  (3,16,4) & 1.5 $\times 10^{-5}$ \\ \hline
  (4,18,5)  & 8.6 $\times 10^{-10}$ \\ \hline
  (6,20,6) & 8.6 $\times 10^{-16}$ \\ \hline
  (8,25,8) & 3.2 $\times 10^{-16}$ \\ \hline
\end{tabular} \quad .
\end{center}

Notice that we chose $\Lambda = 0.5$ to stay sufficiently far away from the poles of the integrand which lie on the imaginary axis. This has the effect of reducing the amplitude of the peaks and the oscillations of the integrand in \eqref{fusionc<=1}, thereby increasing accuracy. 

\begin{figure}[h]
\centering
\begin{subfigure}{0.45\textwidth} 
    \centering
    \includegraphics[scale=0.55]{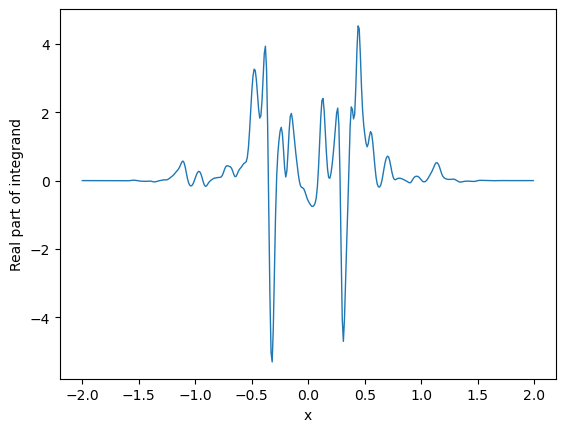} 
    \caption{$\Lambda = 0.05$}
    \label{fig:integrand1}
\end{subfigure}
\hfill 
\begin{subfigure}{0.45\textwidth} 
    \centering
    \includegraphics[scale=0.55]{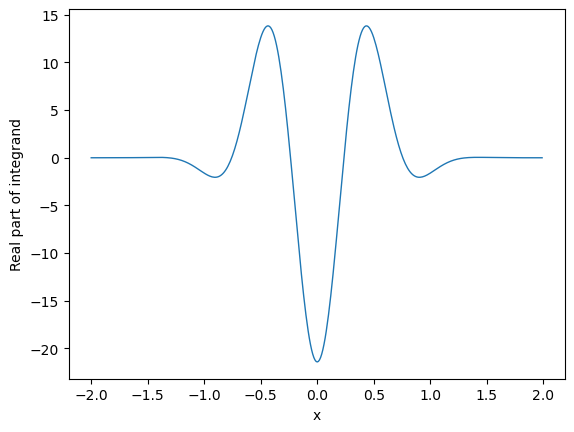} 
    \caption{$\Lambda = 0.5$}
    \label{fig:integrand2}
\end{subfigure}
\caption{Plot of the real part of the integrand in \eqref{fusionc<=1} with $p_t = i x + \Lambda$.}
\label{fig:integrand}
\end{figure}

We were also able to test the fusion transformation \eqref{fusionc<=1}\footnote{In this case we could not use the quad method due to high oscillations of the integrand. We used the method CubicSpline.integrate instead.} for $\Lambda = 0.05$ and with parameters $(T,N_\text{max},L) = (8,20,2)$: the first six digits of the left and right-hand sides agree.  It appears challenging to increase accuracy in this case, because the integrand has rather high oscillations. As an illustration, the real parts of the integrand \eqref{fusionc<=1} for $\Lambda = 0.5$ and $\Lambda = 0.05$ are shown in Figure \ref{fig:integrand}.

Finally, we also verified numerically that the components $\mathbf{\hat F}^+$ and $\mathbf{\hat F}^-$ individually verify the fusion transformation \eqref{fusionc<=1}\footnote{We thank Ioannis Tsiares for suggesting us to perform this check.}. This implies that the integration of the t-channel conformal blocks against $\mathbf{\hat F}^+ - \mathbf{\hat F}^-$ vanishes. This is an important consistency check, since the s-channel conformal blocks are even in $P_s$. 

\subsubsection{Comparing $\mathbf F$ with Ponsot and Teschner's formula}

We also compared the proposal \eqref{defF} and Ponsot-Teschner's formula \eqref{PTformula} for the following values of the parameters:

$$\begin{pmatrix}
b \\ P_1 \\ P_2 \\ P_3 \\ P_4 
\end{pmatrix} = \begin{pmatrix}
 0.77 \\  0.5i \\  0.6i \\ 0.18 i \\  0.31i
\end{pmatrix}, \qquad P_s = 0.35, \quad P_t = 0.65.
$$

In this case, the only parameter which controls the accuracy is the truncation order $T$ of the series. At these values of the parameters, the Ponsot-Teschner formula gives approximatively 0.154 + 2.61$\times 10^{-22}i$. We then performed the following calculations:

\begin{center}
\begin{tabular}{|c|c|c|c|} 
\hline 
  T & $\left|\eqref{defF} - \eqref{PTformula}\right|$ \\ \hline
  0 & 1.7 $\times 10^{-2}$ \\ \hline
  1 & 6.9 $\times 10^{-4}$ \\ \hline
  3 & 2.9 $\times 10^{-6}$ \\ \hline
  5 & 8.3 $\times 10^{-9}$ \\ \hline
  7  & 1.5 $\times 10^{-11}$ \\ \hline
  9 & 2.6 $\times 10^{-14}$ \\ \hline
  12 & 4.5 $\times 10^{-16}$ \\ \hline
\end{tabular} \quad .
\end{center}

\subsection{Relation to earlier works of Ruijsenaars}

The shift operators diagonalized by the fusion (resp. modular) kernels correspond to rank one, quantum relativistic Calogero-Moser Hamiltonians with a hyperbolic potential and associated with the root system $\text{BC}_1$ (resp. $\text{A}_1$) \cite{R2021}. Ruijsenaars' $R$ and $\mathcal R$-functions \cite{R1999R,R2011} (which, up to normalization, are equal to the fusion and modular kernels \cite{R2021}), were constructed as joint eigenfunctions of these Hamiltonians and then interpreted as kernels of a unitary eigenfunction transform \cite{R2003,R2011}. 

Ruijsenaars derived a rigorous series representation for the $\mathcal R$-function in \cite{R2001}, and our proposal \eqref{defM} for $\mathbf M$ is essentially a rewriting of his findings. In particular, he already wrote the series \eqref{seriesg} in \cite[Equations (2.4), (2.59)]{R2001}, however, he admittedly treated it as a formal series and did not study its analytic properties.

Ruijsenaars obtained the series \eqref{seriesg} indirectly by first considering the case 
\begin{equation} \label{P0mn}
P_0 = P_0^{(m,n)}= \frac{iQ}2 - inb - \frac{im}b, \qquad n,m \in \mathbb Z.
\end{equation}
For this special case, he constructed in \cite{R1999} explicit series solutions of the shift equations which, as opposed to \eqref{seriesg}, are finite series. Then, since for $b^2$ irrational the set of values \eqref{P0mn} is dense in the imaginary line, he calculated the interpolation limit in \cite{R2001} and arrived to the formal series \eqref{seriesg}. He also provided similar ideas in the case of the $R$-function in \cite{R2003D4}, however, he did not write explicit formulas due to their higher complexity.

The novelty of the present article as opposed to Ruijsenaars' works is the introduction of the kernels $\hat{\mathbf F}$ and $\hat{\mathbf M}$, and their interpretation as kernels of the fusion and modular transformations for $c<1$. It would be interesting to understand how Ruijsenaars' construction of unitary eigenfunction transforms involving $\mathbf F$ and $\mathbf M$ as kernels generalize to the case of $\hat{\mathbf F}$ and $\hat{\mathbf M}$, and how conformal blocks (which do not appear in his works) fit into this framework.


\subsection{Discussion and outlook} \label{sectiondiscussion}

Below, we mention several other directions that deserve further investigation.

\begin{enumerate}
\item It is important to understand better the analytic properties of the series $f$ \eqref{ansatz}, and to verify that the ones of $\mathbf F$ are inherited from those of $f$. To this end, it would be helpful to relate $f$ to known objects. In fact, it was showed in \cite[Theorem 4.2]{LR} that in the special case $P_s = \frac{iQ}2 + P_1 + P_2 + inb$ where $n \in \mathbb Z_{\geq 0}$, the Virasoro fusion kernel reduces to the celebrated Askey-Wilson polynomials with a quantum deformation parameter $q=e^{2i \pi b^2}$. These polynomials are essentially a basic $q$-hypergeometric series ${}_4 \phi _3$ \cite{KLS2010}.
This leads us to expect that the series
\begin{equation}\label{part_of_f}
\sum_{k=0}^\infty \alpha_k(b,P_s) e^{-2\pi b k P_t}
\end{equation}
can also be related to a basic $q$-hypergeometric series ${}_{n+1} \phi_n$ for some $n$, and where $q = e^{2i \pi b^2}$.  We plan to resort to this question in the future.


\item The limit \eqref{relationFandM} implies that the coefficients $\alpha_n$ in \eqref{alphan} reduce to $\mu_n$ \eqref{mun}. This is highly nontrivial and suggests that there exists a simpler, fully explicit representation for $\alpha_n$ in terms of $q$-Pochhammer symbols. 

\item In \cite{GRSS1}, Ghosal, Remy, Sun and Sun provided a rigorous probabilistic construction of the four-point spherical conformal blocks and proved the fusion transformation \eqref{fusionc>25} for $c>25$ in a certain region of the parameter space. The case of the one-point toric conformal blocks was also proved in \cite{GRSS2}. It would be interesting to obtain our formulas for $\mathbf F$ and $\mathbf M$ within their framework. 

\end{enumerate}

\section{Formal derivation of the result and some properties} \label{derivation}

In this section we show that the proposals \eqref{defF} for $\mathbf F^{(b)}_{P_s,P_t}\left[\substack{P_2\;\;\;P_3\vspace{0.1cm}\\ P_1\;\;P_4}\right]$ and \eqref{defhatF} for $ \mathbf{\hat F}^{(ib)}_{iP_s,iP_t}\left[\substack{iP_2\;\;\;iP_3\vspace{0.1cm}\\ iP_1\;\;iP_4}\right]$ satisfy the same expected shift equations in $P_s$ and $P_t$, and we discuss some of their properties. 

\subsection{Solving the renormalized shift equations} \label{sec2p1}

The Virasoro fusion kernel is known to satisfy two pairs of shift equations in $P_s$ and $P_t$ \cite{R2021}. More precisely, define the shift operators 
\begin{align} 
 \label{HF} & H_{P_s}^{(b)} \left[\substack{P_2\;\;\;P_3\vspace{0.1cm}\\ P_1\;\;P_4}\right] := h(P_s) e^{ib\partial_{P_s}}+h(-P_s) e^{-ib\partial_{P_s}}+ V_{P_s}\left[\substack{P_2\;\;\;P_3\vspace{0.1cm}\\ P_1\;\;P_4}\right], \\
 \label{tildeHF} & \tilde H_{P_t}^{(b)} \left[\substack{P_2\;\;\;P_3\vspace{0.1cm}\\ P_1\;\;P_4}\right] := \tilde h(P_t) e^{ib\partial_{P_t}}+\tilde h(-P_t) e^{-ib\partial_{P_t}} + V_{P_t}\left[\substack{P_2\;\;\;P_1\vspace{0.1cm}\\ P_3\;\;P_4}\right],
\end{align}
with $e^{\pm ib\partial_{P_s}} y(P_s) := y(P_s \pm ib)$ and where
\begin{align*}
& h(P_s)=4\pi^2 \frac{\Gamma \lb 1+2b^2-2ibP_s \rb \Gamma \lb b^2-2i b P_s \rb \Gamma \lb -2i b P_s \rb \Gamma \lb 1+b^2-2ib P_s \rb}{\prod_{\epsilon,\epsilon'=\pm1}\Gamma \lb \tfrac{bQ}2-ib(P_s+\epsilon P_3+\epsilon' P_4)\rb \Gamma \lb \tfrac{bQ}2-ib (P_s+\epsilon P_1+\epsilon' P_2 )\rb}, \\
 & \tilde h(P_t) = 4\pi^2 \frac{~\Gamma \left(1-b^2+2 i bP_t \right) \Gamma (1+2ibP_t) \Gamma \left(2 i b P_t-2 b^2\right) \Gamma \left(2 i b P_t-b^2\right)}{\prod _{\epsilon,\epsilon'=\pm} \Gamma \left(\frac{1-b^2}2+ib \left(P_t+\epsilon P_1+\epsilon' P_4\right)\right) \Gamma \left(\frac{1-b^2}2+ib \left(P_t+\epsilon P_3+\epsilon' P_2\right)\right)},
 \end{align*}
and
\beq\label{H0} \begin{split}
& V_{P_s}\left[\substack{P_2\;\;\;P_3\vspace{0.1cm}\\ P_1\;\;P_4}\right] = -2\operatorname{cosh}{( 2\pi b (P_2+P_3+\tfrac{ib}2))}\\
&+4 \displaystyle \sum_{k=\pm} \frac{\prod_{\epsilon=\pm} \operatorname{cosh}{(\pi b(\epsilon P_4-\tfrac{ib}2-P_3-kP_s))} \operatorname{cosh}{(\pi b(\epsilon P_1-\tfrac{ib}2-P_2-kP_s))}}{\operatorname{sinh}{\lb 2\pi b(kP_s+\frac{ib}2)\rb}\operatorname{sinh}{\lb 2\pi b k P_s \rb}}.
\end{split} \eeq
Then, the Virasoro fusion kernel satisfies \cite{R2021}
\begin{align}
 \label{eqH}   & H_{P_s}^{(b^{\pm1})}\mathbf F^{(b)}_{P_s,P_t} = 2 \cosh(2\pi b^{\pm1} P_t) \mathbf F^{(b)}_{P_s,P_t}, \\
 \label{eqHtilde}   & \tilde H_{P_t}^{(b^{\pm1})}\mathbf F^{(b)}_{P_s,P_t} = 2 \cosh(2\pi b^{\pm1} P_s) \mathbf F^{(b)}_{P_s,P_t}.
\end{align}
Ribault and Tsiares found in \cite{RT} that the image of $\mathbf F$ under the Virasoro-Wick rotation
\begin{equation} \label{Virasoro-Wick}
    \mathcal R \mathbf F^{(b)}_{P_s,P_t}\left[\substack{P_2\;\;\;P_3\vspace{0.1cm}\\ P_1\;\;P_4}\right] := \frac{P_t}{P_s} \mathbf F^{(ib)}_{iP_t,iP_s}\left[\substack{iP_2\;\;\;iP_1\vspace{0.1cm}\\ iP_3\;\;iP_4}\right]
\end{equation}
satisfies the same shift equations. As described in the introduction, the issue is that $\mathcal R \mathbf F$ is odd in $P_s$ and $P_t$, hence it is a nonphysical solution of the shift equations. 

We now show that both the proposals \eqref{defF} and \eqref{defhatF} satisfy the shift equations \eqref{eqH} and \eqref{eqHtilde}. By evenness of the equations in $P_s$, this is equivalent to showing that both $\mathbf F^{+,(b)}_{P_s,P_t}\left[\substack{P_2\;\;\;P_3\vspace{0.1cm}\\ P_1\;\;P_4}\right]$ \eqref{defF+} and $\mathbf{\hat F}^{+,(\beta)}_{p_s,p_t}\left[\substack{p_2\;\;\;p_3\vspace{0.1cm}\\ p_1\;\;p_4}\right]$ \eqref{defhatF+} satisfy \eqref{eqH} and \eqref{eqHtilde}. We start by observing that the following equalities between shift operators hold:
\begin{align}
\label{KD1}  & K_{P_s,P_t}^{(b)} \left[\substack{P_2\;\;\;P_3\vspace{0.1cm}\\ P_1\;\;P_4}\right] D_{P_s}^{(b)}\left[\substack{P_2\;\;\;P_3\vspace{0.1cm}\\ P_1\;\;P_4}\right] \lb K_{P_s,P_t}^{(b)} \left[\substack{P_2\;\;\;P_3\vspace{0.1cm}\\ P_1\;\;P_4}\right]\rb^{-1} = H_{P_s}^{(b)} \left[\substack{P_2\;\;\;P_3\vspace{0.1cm}\\ P_1\;\;P_4}\right], \\
\label{KD2}  & K_{P_s,P_t}^{(b)} \left[\substack{P_2\;\;\;P_3\vspace{0.1cm}\\ P_1\;\;P_4}\right] \tilde D_{P_t}^{(b)} \left[\substack{P_2\;\;\;P_3\vspace{0.1cm}\\ P_1\;\;P_4}\right] \lb K_{P_s,P_t}^{(b)} \left[\substack{P_2\;\;\;P_3\vspace{0.1cm}\\ P_1\;\;P_4}\right]\rb^{-1} = \tilde H_{P_t}^{(b)} \left[\substack{P_2\;\;\;P_3\vspace{0.1cm}\\ P_1\;\;P_4}\right],
\end{align}
as well as
\begin{align}
\label{hatKD1}  & \hat K_{iP_s,iP_t}^{(ib)} \left[\substack{iP_2\;\;\;i P_3\vspace{0.1cm}\\ iP_1\;\;iP_4}\right] D_{P_s}^{(b)}\left[\substack{P_2\;\;\;P_3\vspace{0.1cm}\\ P_1\;\;P_4}\right] \lb \hat K_{iP_s,iP_t}^{(ib)} \left[\substack{iP_2\;\;\;iP_3\vspace{0.1cm}\\ iP_1\;\;iP_4}\right] \rb^{-1} = H_{P_s}^{(b)} \left[\substack{P_2\;\;\;P_3\vspace{0.1cm}\\ P_1\;\;P_4}\right], \\
\label{hatKD2}   & \hat K_{iP_s,iP_t}^{(ib)} \left[\substack{iP_2\;\;\;i P_3\vspace{0.1cm}\\ iP_1\;\;iP_4}\right] \tilde D_{P_t}^{(b)}\left[\substack{P_2\;\;\;P_3\vspace{0.1cm}\\ P_1\;\;P_4}\right] \lb \hat K_{iP_s,iP_t}^{(ib)} \left[\substack{iP_2\;\;\;iP_3\vspace{0.1cm}\\ iP_1\;\;iP_4}\right] \rb^{-1} = \tilde H_{P_t}^{(b)} \left[\substack{P_2\;\;\;P_3\vspace{0.1cm}\\ P_1\;\;P_4}\right],
\end{align}
where $\mathcal D$ and $\tilde{\mathcal D}$ are given by\footnote{The shift operators $\mathcal D$ and $\tilde{\mathcal D}$ were first introduced by Ruijsenaars in \cite{R2003bis}.}
\begin{align}\label{hamiltonianE}
& D_{P_s}^{(b)} \left[\substack{P_2\;\;\;P_3\vspace{0.1cm}\\ P_1\;\;P_4}\right] := e^{-i b \partial_{P_s}}+A^{(b)} _{P_s}\left[\substack{P_2\;\;\;P_3\vspace{0.1cm}\\ P_1\;\;P_4}\right] e^{i b \partial_{P_s}} + V^{(b)} _{P_s}\left[\substack{P_2\;\;\;P_3\vspace{0.1cm}\\ P_1\;\;P_4}\right], \\
\label{hamiltonianEtilde} & \tilde D_{P_t}^{(b)} \left[\substack{P_2\;\;\;P_3\vspace{0.1cm}\\ P_1\;\;P_4}\right] := D_{P_t}^{(b)} \left[\substack{P_2\;\;\;P_1\vspace{0.1cm}\\ P_3\;\;P_4}\right],
\end{align}
with
\begin{align}
 A_{P_s}^{(b)} \left[\substack{P_2\;\;\;P_3\vspace{0.1cm}\\ P_1\;\;P_4}\right] = \frac{16 \cosh(\pi b(P_s \pm P_1 \pm P_2+\frac{ib}2))\cosh(\pi b(P_s \pm P_3 \pm P_4+\frac{ib}2))}{\sinh(2\pi b P_s) \sinh(2\pi b(P_s + \frac{ib}2))^2 \sinh(2\pi b(P_s + ib))}.
\end{align}
These identities can be verified utilizing the shift identities for the double Gamma function
\begin{equation} \label{shiftdoublegamma}
    \frac{\Gamma_b(z+b)}{\Gamma_b(z)} = \frac{\sqrt{2\pi} \; b^{bz-\frac12}}{\Gamma(bz)}, \qquad (b \to b^{-1}).
\end{equation}
Next, \eqref{KD1} and \eqref{KD2} can be used to rewrite the shift equations \eqref{eqH} and \eqref{eqHtilde} for $\mathbf F^{+,(b)}_{P_s,P_t}\left[\substack{P_2\;\;\;P_3\vspace{0.1cm}\\ P_1\;\;P_4}\right]$ in terms of shift equations for the series $f$:
\begin{align}
  \label{eqDPsf} & D_{P_s}^{(b^{\pm1})}\left[\substack{P_2\;\;\;P_3\vspace{0.1cm}\\ P_1\;\;P_4}\right] f_{P_s,P_t}^{(b)}\left[\substack{P_2\;\;\;P_3\vspace{0.1cm}\\ P_1\;\;P_4}\right]=2 \cosh(2\pi b^{\pm1} P_t) f_{P_s,P_t}^{(b)}\left[\substack{P_2\;\;\;P_3\vspace{0.1cm}\\ P_1\;\;P_4}\right], \\
   \label{eqDPtf} & \tilde D_{P_t}^{(b^{\pm1})}\left[\substack{P_2\;\;\;P_3\vspace{0.1cm}\\ P_1\;\;P_4}\right] f_{P_s,P_t}^{(b)}\left[\substack{P_2\;\;\;P_3\vspace{0.1cm}\\ P_1\;\;P_4}\right] = 2 \cosh(2\pi b^{\pm1} P_s) f_{P_s,P_t}^{(b)}\left[\substack{P_2\;\;\;P_3\vspace{0.1cm}\\ P_1\;\;P_4}\right].
\end{align}
Similarly, \eqref{hatKD1} and \eqref{hatKD2} imply that \eqref{eqH} and \eqref{eqHtilde} for $\mathbf{\hat F}^{+,(\beta)}_{p_s,p_t}\left[\substack{p_2\;\;\;p_3\vspace{0.1cm}\\ p_1\;\;p_4}\right]$ are equivalent to 
\begin{align}
  \label{eqDPsfimag} & D_{P_s}^{(b^{\pm1})}\left[\substack{P_2\;\;\;P_3\vspace{0.1cm}\\ P_1\;\;P_4}\right] f_{iP_s,iP_t}^{(ib)} \left[\substack{iP_2\;\;\;iP_3\vspace{0.1cm}\\ iP_1\;\;iP_4}\right]=2 \cosh(2\pi b^{\pm1} P_t) f_{iP_s,iP_t}^{(ib)} \left[\substack{iP_2\;\;\;iP_3\vspace{0.1cm}\\ iP_1\;\;iP_4}\right] \\
   \label{eqDPtfimag} & \tilde D_{P_t}^{(b^{\pm1})}\left[\substack{P_2\;\;\;P_3\vspace{0.1cm}\\ P_1\;\;P_4}\right] f_{iP_s,iP_t}^{(ib)} \left[\substack{iP_2\;\;\;iP_3\vspace{0.1cm}\\ iP_1\;\;iP_4}\right] = 2 \cosh(2\pi b^{\pm1} P_s) f_{iP_s,iP_t}^{(ib)} \left[\substack{iP_2\;\;\;iP_3\vspace{0.1cm}\\ iP_1\;\;iP_4}\right].
\end{align}
It is now readily seen that \eqref{eqDPsf} and \eqref{eqDPtf} imply \eqref{eqDPsfimag} and \eqref{eqDPtfimag}, because the shift operators $\mathcal D$ and $\tilde{\mathcal D}$ are invariant under the transformations $b \to ib$ and $P_j \to i P_j$, $j=1,2,3,4,s,t$. 
It then remains to show that \eqref{eqDPsf} and \eqref{eqDPtf} hold. Notice from \eqref{alphan} that $e^{\pm i b \partial_{P_s}} \alpha_l(b^{-1},P_s) = \alpha_l(b^{-1},P_s)$. Then, substitution of \eqref{ansatz} into \eqref{eqDPsf} with the plus sign leads to 
\begin{align*}
e^{2i\pi P_s P_t} \sum_{k=0}^\infty \sum_{l=0}^\infty & \alpha_l(b^{-1},P_s) \bigg[ \alpha_{k+1}(b,P_s-ib) + \alpha_{k-1}(b,P_s+ib) A_{P_s}^{(b)} \\
& + \alpha_k(b,P_s) V_{P_s}^{(b)} - \alpha_{k-1}(b,P_s) - \alpha_{k+1}(b,P_s) \bigg] e^{-2\pi k b P_t} e^{-\frac{2\pi l P_t}{b}} = 0.
\end{align*}
We now claim that the family of coefficients $\alpha_n$ defined in \eqref{alphan} is the unique solution of the shift-recurrence relation
\begin{equation} \label{shiftrecurrence}
    \alpha_{k+1}(b,P_s-ib) + \alpha_{k-1}(b,P_s+ib) A_{P_s}^{(b)} + \alpha_k(b,P_s) V_{P_s}^{(b)} - \alpha_{k-1}(b,P_s) - \alpha_{k+1}(b,P_s) = 0
\end{equation}
which satisfies $\alpha_n(b,P_s) \to 0$ as $\re(P_s) \to +\infty$. We conclude that the series f satisfies the two shift equations \eqref{eqDPsf}, since $f$ is invariant under $b\to b^{-1}$. Finally, we verified numerically in the domain of convergence of the series that 
    \begin{equation} \label{selfdual}
      f_{P_s,P_t}^{(b)} \left[\substack{P_2\;\;\;P_3\vspace{0.1cm}\\ P_1\;\;P_4}\right] = f_{P_t,P_s}^{(b)} \left[\substack{P_2\;\;\;P_1\vspace{0.1cm}\\ P_3\;\;P_4}\right].
    \end{equation}
Therefore, $f$ also verifies \eqref{eqDPtf}. Let us remark that we do not have an analytic proof that $\alpha_n$ in \eqref{alphan} satisfies \eqref{shiftrecurrence}, however, we verified it numerically up to $k=50$. 


Finally, Ruijsenaars rigorously proved in \cite{R2003bis} the leading asymptotics of of the $R$-function (which is essentially $\mathbf F$ \cite{R2021}) as $\re P_t \to +\infty$. By evenness in $P_t$, the case $\re P_t \to -\infty$ obviously follows. We added a phase in \eqref{definitionKc>1} (which does not affect the shift equations) so that our formula \eqref{defF} reduces exactly to Ruijsenaars' asymptotic result when only the first term in the series $f$ is kept. We believe this requirement eliminates the need to study the shift equations in the external momenta satisfied by the fusion kernel \cite{E}. Moreover, we added a similar phase and a factor -i in \eqref{definitionKc<=1} so that our proposals behave under Virasoro-Wick rotations just like in \eqref{Virasoro-Wick}. 

\subsection{The Virasoro-Wick rotation}

Our proposals \eqref{defF} and \eqref{defhatF} for the Virasoro fusion kernels at $c > 25$ and $c < 1$ resemble each other. In fact, the components $\mathbf F^\pm$ and $\mathbf{\hat F}^\pm$ are related by a Virasoro-Wick rotation \eqref{Virasoro-Wick}. More precisely, in this section we show that for $j=\pm$ we have
\begin{equation}
    \mathcal R \mathbf{F}^{j,(b)}_{P_s,P_t}\left[\substack{P_2\;\;\;P_3\vspace{0.1cm}\\ P_1\;\;P_4}\right] = -j i \mathbf{\hat F}^{j,(\beta)}_{p_s,p_t}\left[\substack{p_2\;\;\;p_3\vspace{0.1cm}\\ p_1\;\;p_4}\right].
\end{equation}
The case $j=+$ is straightforward, because thanks to the identity \eqref{selfdual} the two series cancel each other. It then suffices to handle the ratios of $\Gamma_{ib}$ functions by using the shift identity \eqref{shiftdoublegamma}.
The case $j=-$ necessitates an extra step. More precisely, we have
\begin{equation}
\frac{\mathcal R \mathbf{F}^{-,(b)}_{P_s,P_t}\left[\substack{P_2\;\;\;P_3\vspace{0.1cm}\\ P_1\;\;P_4}\right]}{i \mathbf{\hat F}^{-,(\beta)}_{p_s,p_t}\left[\substack{p_2\;\;\;p_3\vspace{0.1cm}\\ p_1\;\;p_4}\right]} = \frac{p_t}{p_s} \frac{K_{-p_t,p_s}^{(\beta)} \left[\substack{p_2\;\;\;p_1\vspace{0.1cm}\\ p_3\;\;p_4}\right]}{\hat K_{-p_s,p_t}^{(\beta)} \left[\substack{p_2\;\;\;p_3\vspace{0.1cm}\\ p_1\;\;p_4}\right]} \frac{f_{-p_t,p_s}^{(\beta)} \left[\substack{p_2\;\;\;p_1\vspace{0.1cm}\\ p_3\;\;p_4}\right]}{f_{-p_s,p_t}^{(\beta)} \left[\substack{p_2\;\;\;p_3\vspace{0.1cm}\\ p_1\;\;p_4}\right]}.
\end{equation}
We then use the identity \eqref{secondidentityf} as well as the definition $S_\beta(z) = \Gamma_\beta(z) / \Gamma_\beta(\hat Q-z)$. The remaining part of the computation is straightforward: the resulting ratio of functions $\Gamma_\beta$ simplifies nicely thanks to \eqref{shiftdoublegamma}.

\subsection{Domain of convergence of the series $f$} 

We now formulate a conjecture regarding the convergence of the series \eqref{part_of_f}. A similar conjecture can be formulated for the series entering the modular kernel.

\begin{cj} \label{cj1}
Let $b^2 \notin \mathbb Q$ and such that $b^2$ is not a Liouville number, that is,
\begin{equation}
\exists m, q_0 \in \mathbb N, \qquad \forall (p,q) \in \mathbb Z \times \mathbb Z_{\geq q_0}, \qquad \left| b^2 - \frac{p}{q} \right| > \frac{1}{q^m}.
\end{equation}
Then, the series \eqref{part_of_f} has a positive radius of convergence whose value depends on $b, P_s,P_1,P_2,P_3,P_4$.
\end{cj}
The motivation for this conjecture originates from the expectation that the series \eqref{part_of_f} can be related to a basic $q$-hypergeometric series with $q=e^{2i\pi b^2}$ (see the beginning of Section \ref{sectiondiscussion}). In fact, it was showed in \cite{O} that if $b^2$ is irrational and not a Liouville number, then basic $q$-hypergeometric series of type ${}_{n+1} \phi_n$ with $|q| = |e^{2i\pi b^2}| =1$ have a positive radius of convergence which depends on its parameters. If such conditions on $b$ are not met, ${}_{n+1} \phi_n$ may diverge. Finally, we notice that the assumptions on $b^2$ in Conjecture \ref{cj1} appear natural because, recalling the definition \eqref{alphan} of the coefficients $\alpha_n$, the assumption that $b^2 \notin \mathbb Q$ implies that $\sin(n \pi b^2) \neq 0$, whereas the assumption that $b^2$ is not a Liouville number implies that $\tfrac{1}{\sin(n \pi b^2)}$ admits a polynomial bound as $n \to \infty$ \cite{Rib20}\footnote{We thank Sylvain Ribault for suggesting the importance of $b^2$ not being a Liouville number.}.

The assumption on $b^2$ cannot be realized numerically. Instead, we conducted numerical tests of convergence by choosing $b^2 = M/N$ with $M$ and $N$ coprime integers and with $M, N$ sufficiently large.  We could always find a pair $(b,P_t)$ such that $|\text{exp}(-2\pi b P_t)|$ is small enough and so that the series converges quickly. However, whenever $|\text{exp}(-2\pi b P_t)| \geq 1$ the series does not seem to converge. This suggests that the radius of convergence of the series is always smaller than 1. In particular, one assumption which seems necessary for convergence is that $\Re(P_t) >0$. 

In the cases $\beta \in \mathbb R$ and $p_t, p_s \in i \mathbb R$ (which are relevant for Liouville theory with $c\leq 1$ \cite{RS}), we have $|\text{exp}(-2\pi \beta^{\pm 1}p_t)| = 1$, and the series $f_{p_s,p_t}^{(\beta)} \left[\substack{p_2\;\;\;p_3\vspace{0.1cm}\\ p_1\;\;p_4}\right]$ apparently diverges. However, in the fusion transformation \eqref{fusionc<=1} a nonzero real value $\Lambda$ is added to $p_t$. We can choose it to be positive and large enough so that the series converges. \\

\subsubsection*{Ribault and Tsiares' conjecture \cite{RT}} 

On the first hand, if Conjecture \ref{cj1} is true, it is likely to imply that there exists a region in the parameter space for which the series $f$ is absolutely convergent and meromorphic. On the other hand, Ribault and Tsiares conjectured in \cite{RT} that the physical fusion kernel for $c<1$ is a distribution.  This seems to lead to a contradiction. However, we believe that these two facts are consistent with each other for the following reason (see also \cite{RT} for a similar discussion). 

The relation between our series representation \eqref{defF} for $\mathbf F$ and Ruijsenaars' asymptotic result \cite{R2003bis} (see Section \ref{sec2p1}) strongly suggests that \eqref{defF} should be understood as a complete asymptotic expansion of the Ponsot-Teschner formula \eqref{PTformula} as $\re P_t > \delta$ and for some sufficiently large $\delta$. By analogy with the case of $\mathbf F$, it could well be that there exists a distribution which is manifestly even in $P_t$ and whose asymptotic expansion for $\Re(P_t) > \delta$ corresponds to \eqref{defhatF}. 

\subsubsection*{Uniqueness of the solutions of the shift equations for $c \notin \mathbb Q$ and $c \in \mathbb C \backslash (-\infty,1]$}

On the first hand, for $c \notin \mathbb Q$ and $c \in \mathbb C \backslash (-\infty,1]$ the solution space of the shift equations consisting of meromorphic functions forms a one-dimensional vector space over $\mathbb C$ and is spanned by $\mathbf F$ \cite{E}. On the other hand, we observed that both the components $\mathbf F^+$ and $\mathbf F^-$ satisfy the shift equations and can be used as integral kernels in \eqref{fusionc>25}. This observation does not contradict the results of \cite{E}, since $\mathbf F^\pm$ do not belong to the solution space of \cite{E}. Indeed, while $\mathbf F^\pm$ are certainly not meromorphic for $\Re(P_t)<0$, one crucial assumption of \cite{E} is that, as functions of $P_t$, the solutions are meromorphic across the entire complex plane. 

\subsection{The limit $b^2$ rational}

If $b^2 = \frac{M}{N}$ with $M$ and $N$ coprime integers, then it can be seen from \eqref{phinn} that the coefficients $\phi_{nn}$ (and, therefore, $\alpha_n(b,P_s)$) diverge for $n\geq N$. Similarly, in this case we also have that $\alpha_n(b^{-1},P_s)$ diverges for $n \geq M$. However, rather surprisingly, it might well be that the series $f$ itself has a well-defined limit as $b^2 \to \frac{M}{N}$ \footnote{A similar observation was made by Ruijsenaars in the case of the series $g$ entering $\mathbf M$ \cite{R2001}.}. For instance, in the case $b^2 = 1$ we observed that the first two terms in the series, namely,
\begin{align*}
& \alpha_1(b,P_s) e^{-2\pi b P_t} + \alpha_1(b^{-1},P_s) e^{-2\pi b^{-1} P_t}, \\
& \alpha_2(b,P_s) e^{-4\pi b P_t} + \alpha_2(b^{-1},P_s) e^{-4\pi b^{-1} P_t} + \alpha_1(b,P_s) \alpha_1(b^{-1},P_s) e^{-2\pi b P_t}e^{-2\pi b^{-1} P_t},
\end{align*}
have a well-defined limit as $b\to 1$.  Moreover, the numerical tests of the fusion transformation \eqref{fusionc<=1} were performed for $\beta^2 = \frac{9}{25}$. This indicate that the apparent divergence of the series $f$ at rational values of $\beta^2$ is merely an artifact of the formulas, and the limit should be well-defined.

We then conjecture that the limit $\beta^2$ rational of $f$ is well-defined. If true,  it would mean that the limits of $\mathbf{F}$ as $c \to 25$ and of $\mathbf{\hat{F}}$ as $c \to 1$ are well-defined. It would be interesting to compare such limits with the known fusion kernels. At $c=1$, it is proportional to the connection constant for the Painlev\'e VI tau function \cite{GIL,ILT}, whereas at $c=25$ a non-integral representation was recently constructed by Ribault and Tsiares in \cite{RT}.

\subsection{$W(D_4)$ invariance of $f$}

It was observed in \cite{LNR} that the Virasoro conformal blocks possess discrete symmetries under three flips $s_i: P_i \to - P_i$ for $i=1,2,3$ and under the so-called Regge-Okamoto transformation $s_\delta: P_i \to P_i - \delta$ with $\delta = \frac12 \sum_{i=1}^4 P_i$. These four transformations generate the Weyl group $W(D_4)$ of the Lie algebra of type $D_4$ (notice that the fourth flip $s_4$ is a certain product of the generators \cite{LNR}). 

It can readily be observed that the shift operators $D$ and $\tilde D$ in \eqref{hamiltonianE} and \eqref{hamiltonianEtilde} -- which admit the series $f$ as a joint eigenfunction -- are $W(D_4)$-invariant (this was noticed first by Ruijsenaars in \cite{R2003bis}). Although this does not necessarily imply that $f$ also has this invariance, we expect it to be true. In fact,  the coefficient $\alpha_1$ in \eqref{alpha1} is clearly $W(D_4)$-invariant, and we verified numerically that all $\alpha_n$ up to high order $n$ also have this invariance.

\section{Consistency checks}

\subsection{The cases $c=1$ and $c=25$ with special external momenta}

We now show that for special external momenta, namely $\Delta(P_i) = \frac{15}{16}$ (resp. $\Delta(P_i) = \frac{1}{16}$), the fusion kernels $\mathbf{F}$ for $c \to 25$ (resp. $\mathbf{\hat{F}}$ for $c \to 1$) have well defined limits which correspond to the known formulas \cite{RT}. More precisely, we have
\begin{align}
   \label{casec=25} & \lim_{b \to 1} \mathbf F^{\pm, (b)}_{P_s,P_t}\left[\substack{\frac{ib}4 \;\;\; \frac{ib}4 \vspace{0.1cm}\\ \frac{ib}4 \;\; \frac{i}{2b}-\frac{ib}4}\right] = \mp i \frac{P_t}{P_s} 16^{P_s^2-P_t^2} e^{\pm 2i\pi P_s P_t}, \\
   \label{casec=1} & \lim_{\beta \to 1} \mathbf{\hat F}^{\pm, (\beta)}_{p_s,p_t}\left[\substack{\frac{i\beta}4\;\;\; \frac{i\beta}4 \vspace{0.1cm}\\ \frac{i\beta}4 \;\; \frac{i}{2\beta} - \frac{i\beta}4}\right] = 16^{p_t^2-p_s^2} e^{\pm 2i\pi p_s p_t},
\end{align}
which implies that
\begin{align}
   \label{casec=25} & \lim_{b \to 1} \mathbf F^{(b)}_{P_s,P_t}\left[\substack{\frac{ib}4 \;\;\; \frac{ib}4 \vspace{0.1cm}\\ \frac{ib}4 \;\; \frac{ib}4 - \frac{i}{2b}}\right] = \frac{P_t}{P_s} 16^{P_s^2-P_t^2} \sin(2\pi P_s P_t), \\
   \label{casec=1} & \lim_{\beta \to 1} \mathbf F^{(\beta)}_{p_s,p_t}\left[\substack{\frac{i\beta}4\;\;\; \frac{i\beta}4 \vspace{0.1cm}\\ \frac{i\beta}4 \;\; \frac{i}{2\beta} - \frac{i\beta}4} \right] = 16^{p_t^2-p_s^2} \cos(2\pi p_s p_t).
\end{align}
The cases $c=25$ and $c=1$ are proved in the same way, hence let us focus on $c=25$. Notice that it is a priori puzzling that we do not send all $P_i$'s to $ib/4$. The key is that by continuity, it should not matter how we approach the values $P_i = i/4$. However, we observed that the way \eqref{casec=25} approaches the limit makes the computation trivial, as we see below. 

We consider the limit \eqref{casec=25} of $\mathbf F^{+, (b)}$, since the case of $\mathbf F^{-, (b)}$ consists of sending $P_s \to -P_s$. The crucial observation is that all coefficients $\phi_{nn}(P_1,P_2,P_3,P_4)$ and $\phi_{nn}(P_1,P_2+\frac{i}{2b},P_3+\frac{i}{2b},P_4)$, as well as their $b \to b^{-1}$ counterparts, vanish for all $n>0$ at $P_1=P_2=P_3 = \frac{ib}4$ and $P_4 = \frac{ib}4 - \frac{i}{2b}$. This implies that at these values $\alpha_{>0}(b,P_s) = \alpha_{>0}(b^{-1},Ps) = 0$. Therefore, only the terms $k=l=0$ in the sum \eqref{ansatz} are nonzero and we obtain
\begin{equation}
    f_{P_s,P_t}^{(b)}\left[\substack{\frac{ib}4 \;\;\; \frac{ib}4 \vspace{0.1cm}\\ \frac{ib}4 \;\; \frac{ib}4 - \frac{i}{2b}}\right] = e^{2i\pi P_s P_t}.
\end{equation}
It remains to compute the limit of the prefactor $K$. Utilizing the identity
\begin{equation}
    \Gamma_1(z) = \frac{(2\pi)^{\frac{z-1}2}}{G(z)}
\end{equation}
where $G(z)$ is the Barnes G-function, we obtain
\begin{align*}
& \lim_{b\to 1} K_{P_s,P_t}^{(b)}\left[\substack{\frac{ib}4 \;\;\; \frac{ib}4 \vspace{0.1cm}\\ \frac{ib}4 \;\; \frac{ib}4 - \frac{i}{2b}}\right] \\
& = i (2\pi)^{-2i(P_s+P_t)} \frac{G(2-2iP_t) G(-2iP_t)}{G(2+2iP_s)G(2iP_s)} \frac{G(\frac12 + i P_s)^2 G(1+iP_s)^4 G(\frac32+iP_s)^2}{G(\frac12-iP_t)^2 G(1-iP_t)^4 G(\frac32-iP_t)^2}.
\end{align*}
Thanks to the doubling identity 
\begin{equation}
    G(2z) = C \; 2^{2z(z-1)} (2\pi)^{-z} G(\frac12 +z)^2 G(1+z) G(z)
\end{equation}
where $C$ is an unimportant constant, as well as $G(1+z) = \Gamma(z) G(z)$ and $\Gamma(1+z) = z \Gamma(z)$, we arrive at
\begin{align}
\lim_{b\to 1} K_{P_s,P_t}^{(b)}\left[\substack{\frac{ib}4 \;\;\; \frac{ib}4 \vspace{0.1cm}\\ \frac{ib}4 \;\; \frac{ib}4 - \frac{i}{2b}}\right] & = i 16^{P_s^2-P_t^2} \frac{G(1+iP_s)^2}{G(2+iP_s)G(iP_s)} \frac{G(2-iP_t) G(-iP_t)}{G(1-iP_t)^2} \\
\nonumber & = i 16^{P_s^2-P_t^2} \frac{\Gamma(iP_s)}{\Gamma(1+iP_s)} \frac{\Gamma(1-iP_t)}{\Gamma(-iP_t)} \\
\nonumber & = -i 16^{P_s^2-P_t^2} \frac{P_t}{P_s}
\end{align}
which leads to the desired result.

\subsection{Crossing symmetry of Liouville theory}

In this section we show that each component $\mathbf F^\pm$ and $\mathbf{\hat F}^\pm$ individually satisfies the crossing symmetry equations of Liouville theory on the four-point Riemann sphere. This is perhaps not surprising, since we verified numerically that each component individually satisfies the fusion transformation.

There exists a normalization of the primary fields in which the two and three-point correlation functions of Liouville theory for $c \in \mathbb C \backslash (-\infty,1]$ are \cite{Rib}
\begin{equation} \label{DOZZ}
    B_P^{b)} = \prod_{\pm} \Gamma_b(\pm 2i P) \Gamma_b(Q\pm 2iP), \qquad C^{(b)}_{P_1,P_2,P_3} = \prod_{\pm, \pm, \pm} \Gamma_b(\tfrac{Q}2\pm iP_1 \pm iP_2 \pm iP_3).
\end{equation}
Similarly, for $c \leq 1$ we have
\begin{equation} \label{imagDOZZ}
    \hat B_P^{b)} = \frac{1}{4P^2 B_{iP}^{(ib)}}, \qquad \hat C^{(b)}_{P_1,P_2,P_3} = \frac{1}{C^{(ib)}_{iP_1,iP_2,iP_3}}.
\end{equation}
The crossing symmetry equations for the four-point correlation function on the sphere for $c \in \mathbb C \backslash (-\infty,1]$ and $c \leq 1$ can then be recasted in terms of the fusion kernels:
\begin{align}
  \label{crossing>1}    & \frac{C^{(b)}_{P_1,P_2,P_s} C^{(b)}_{P_s,P_3,P_4}}{B_{P_s}^{(b)}} \mathbf F^{(b)}_{P_s,P_t}\left[\substack{P_2\;\;\;P_3\vspace{0.1cm}\\ P_1\;\;P_4}\right] = \frac{C^{(b)}_{P_2,P_3,P_t} C^{(b)}_{P_1,P_4,P_t}}{B_{P_t}^{(b)}} \mathbf F^{(b)}_{P_t,P_s}\left[\substack{P_2\;\;\;P_1\vspace{0.1cm}\\ P_3\;\;P_4}\right], \\
 \label{crossing<1} & \frac{\hat C^{(b)}_{P_1,P_2,P_s} \hat C^{(b)}_{P_s,P_3,P_4}}{\hat B_{P_s}^{(b)}} \mathbf{\hat F}^{(b)}_{P_s,P_t}\left[\substack{P_2\;\;\;P_3\vspace{0.1cm}\\ P_1\;\;P_4}\right] = \frac{\hat C^{(b)}_{P_2,P_3,P_t} \hat C^{(b)}_{P_1,P_4,P_t}}{\hat B_{P_t}^{(b)}} \mathbf{\hat F}^{(b)}_{P_t,P_s}\left[\substack{P_2\;\;\;P_1\vspace{0.1cm}\\ P_3\;\;P_4}\right].
\end{align}
Thanks to the duality \eqref{selfdual}, it can readily be verified that $\mathbf F^+$ and $\mathbf{\hat F}^+$ satisfy \eqref{crossing>1} and \eqref{crossing<1}, respectively (note that for the latter we also need the shift identity \eqref{shiftdoublegamma}). It then remains to show that $\mathbf F^-$ and $\mathbf{\hat F}^-$ satisfy the same equations. However, in this case the series do not trivially cancel. Keeping track of all factors, we find that $\mathbf F^-$ and $\mathbf{\hat F}^-$ respectively satisfy \eqref{crossing>1} and \eqref{crossing<1} if the series $f$ satisfies
\begin{equation} \label{secondidentityf}
    \frac{f_{-P_s,P_t}^{(b)} \left[\substack{P_2\;\;\;P_3\vspace{0.1cm}\\ P_1\;\;P_4}\right]}{f_{-P_t,P_s}^{(b)} \left[\substack{P_2\;\;\;P_1\vspace{0.1cm}\\ P_3\;\;P_4}\right]} = u^{(b)}_{P_s}\left[\substack{P_2\;\;\;P_3\vspace{0.1cm}\\ P_1\;\;P_4}\right] u^{(b)}_{-P_t}\left[\substack{P_2\;\;\;P_1\vspace{0.1cm}\\ P_3\;\;P_4}\right],
\end{equation}
where
\begin{equation*}
    u^{(b)}_{Ps}\left[\substack{P_2\;\;\;P_3\vspace{0.1cm}\\ P_1\;\;P_4}\right] = \frac{S_b(Q+2iP_s)}{S_b(Q-2iP_s)} \frac{S_b(\frac{Q}2-iP_1\pm iP_2-iP_s) S_b(\frac{Q}2-iP_3\pm i P_4-iP_s)}{S_b(\frac{Q}2-iP_1\pm iP_2+iP_s) S_b(\frac{Q}2-iP_3\pm i P_4+iP_s)}.
\end{equation*}
We verified this identity numerically up to high order in the domain of convergence of the series. Let us finally mention that an identity similar to \eqref{secondidentityf} in the case of the one-point torus satisfied by the series $g$ \eqref{seriesg} was proved by Ruijsenaars, see \cite[Equation (2.29)]{R2001}.

\subsection{Pairwise identical external operators with exchange of the identity in the S-channel}

In the case of pairwise identical external operators with exchange of the identity in the S-channel, that is, $P_1 = P_2$, $P_3 = P_4$ and $P_s = \frac{iQ}2$, it is well-known \cite{CMMT,E} that the Virasoro fusion kernel reduces to a renormalized version of the structure constants \eqref{DOZZ} of Liouville theory with $c>25$. More precisely, denoting
\begin{align}
& C_0(P_1,P_2,P_3) := \frac{1}{\sqrt{2}} \frac{\Gamma_b(2Q)}{\Gamma_b(Q)^3} \frac{ C^{(b)}_{P_1,P_2,P_3}}{\prod_{j=1}^3 \Gamma_b(Q \pm 2iP_j)}, \\
& \rho_0(P):= \frac{\sqrt2}{S_b(2iP)S_b(-2iP)} = 4 \sqrt{2} \sinh(2\pi b P) \sinh(2\pi b^{-1}P),
\end{align}
We have\footnote{There is a factor $2$ in the left-hand side because we defined $\mathbf F$ as the kernel of the fusion transformation \eqref{fusionc>25} where the integral is taken on $\mathbb R$ and not $\mathbb R_+$.}
\begin{equation} \label{pairwise}
2 \mathbf F^{(b)}_{\frac{iQ}2,P_t}\left[\substack{P_1\;\;\;P_4\vspace{0.1cm}\\ P_1\;\;P_4}\right] = C_0(P_1,P_4,P_t) \rho_0(P_t).
\end{equation}
We verified numerically that our proposal satisfies this identity. More specifically, the series $f$ 
entering $\mathbf F^\pm$ appear to be well-defined in this limit, however, $\mathbf F^+$ vanishes because the prefactor $K$ has a zero coming from $\Gamma_b$-functions. Therefore, our proposal satisfies \eqref{pairwise} provided that
\begin{equation}
\mathbf F^{(b),-}_{\frac{iQ}2,P_t}\left[\substack{P_1\;\;\;P_4\vspace{0.1cm}\\ P_1\;\;P_4}\right] = C_0(P_1,P_4,P_t) \rho_0(P_t).
\end{equation}
This boils down to yet another mysterious identity satisfied by $f$
\begin{equation}
f_{-\frac{iQ}2,P_t}^{(b)} \left[\substack{P_1\;\;\;P_4\vspace{0.1cm}\\ P_1\;\;P_4}\right] = e^{-2i\pi(\frac{Q^2-1}8+P_1^2+P_4^2)} S_b(Q-2iP_t) S_b\lb \tfrac{Q}2 \pm iP_1 \pm iP_4 + iP_t\rb
\end{equation}
that we verified numerically in its domain of convergence.

\subsection{The modular S matrix for the Virasoro characters}


When the external field is the identity field, that is $P_0 = \frac{iQ}2$, the Virasoro modular kernel $\mathbf M$ becomes proportional to the modular S matrix for the Virasoro characters \cite{CMMT}. We now verify that the formula \eqref{defM} satisfies this limit. In fact, when $P_0 = \frac{iQ}2$ (that is, $\alpha_0=0$) the coefficients $\mu$ in \eqref{mun} satisfy $\mu_{\geq 1} = 0$. A straightforward computation then shows that
\begin{equation}
    \mathbf M^{\pm,(b)}_{P_s,P_t}\left[\frac{iQ}2\right] = \sqrt{2} \; e^{\pm 4i \pi P_s P_t},
\end{equation}
hence we have
\begin{equation}
    \mathbf M^{(b)}_{P_s,P_t}\left[\frac{iQ}2\right] = \sqrt{2} \; \cos(4\pi P_s P_t),
\end{equation}
as expected from \cite{CMMT}. 

\section*{Acknowledgements}

We are grateful to thank Sylvain Ribault for illuminating discussions, and for his helpful comments on an earlier version of the draft and on the ancillary Jupyter notebook. We also thank Simon Ruijsenaars and Ioannis Tsiares for helpful discussions. We finally thank an anonymous referee for useful suggestions. We are supported by the Academy of Finland Centre of Excellence Programme grant number 346315 entitled "Finnish centre of excellence in Randomness and STructures (FiRST)".

\end{document}